# Mastering Preclinical Fast MRSI: From Setup to Execution

Gianna Nossa[1,2,†,c], Eloïse Mougel[1,2,†,c], Brayan Alves[1,2], Tan Toi Phan[1,2], Alessio Siviglia[1,2], Thi Ngoc Anh Dinh[1,2], Thanh Phong Lê[1,2], Bernard Lanz[1,2], Cristina Cudalbu[1,2]

*Affiliations*
[1]CIBM Center for Biomedical Imaging, Lausanne, Switzerland
[2]CIBM Pre-Clinical Imaging EPFL Metabolic Imaging Section, EPFL - École polytechnique fédérale de Lausanne, Lausanne, Switzerland

† This author contributed equally to this work

c. corresponding author:
Name: Gianna Nossa
email address: gianna.nossa@epfl.ch
telephone: +41 21 693 51 44

c. corresponding author:
Name: Eloïse Mougel
email address: eloise.mougel@epfl.ch
telephone: +41 21 693 68 48

**Abstract (200 words max)**

MR experiments are essential for studying brain metabolism, yet preclinical $^1$H-MRSI remains underdeveloped, with significant limitations in SNR, acquisition speed, and automated data processing. Although recent advances—such as accelerated sequences, denoising strategies, and ultra-high-field systems—have begun to reduce these barriers, preclinical MRSI still lags far behind the human research field in accessibility and routine use. Based on our expertise, we have created this guide that outlines a complete workflow for acquiring and analyzing high-quality fast MRSI data in rodent brains at ultra-high fields (9.4T and 14.1T), enabling novice users to perform reliable experiments using optimized MRSI sequences (FID-MRSI, SE-MRSI, and PRESS-MRSI) and standardized processing pipelines, while also highlighting strategies to further improve acquisition speed, coverage, and reproducibility. Overall, this paper provides a strong foundation for future methodological advances that will expand metabolic imaging capabilities and deepen insights into brain function and disease.

## 1. Introduction

In experimental research, rodent models have long played a central role in investigating brain development, degeneration, function and metabolism, elucidating pathological processes and guiding therapeutic innovation. Understanding cerebral metabolism requires a multiscale approach that captures interactions across different brain regions simultaneously. $^1$H spectroscopic imaging ($^1$H-MRSI) has emerged as a key technique in this context, enabling spatially resolved metabolite mapping and offering unique opportunities to probe regional metabolism [1, 2]. Research on $^1$H-MRSI using clinical scanners is well advanced and has benefited from the recent democratisation of high-field systems, as well as sophisticated acquisition and reconstruction methodologies [1, 3, 4]. These ongoing developments are beginning to demonstrate the substantial clinical value of $^1$H-MRSI across multiple applications, including diagnosis, monitoring, and treatment planning [1], sending this modality on a clear trajectory toward broader clinical adoption. In contrast, preclinical applications remain in earlier stages of development, creating a marked difference in the level of implementation between human and rodent imaging.

Translating $^1$H-MRSI to small-animal imaging introduces several challenges. The markedly smaller brain volume of rodents, typically between 100 and 500 µL, necessitates high spatial resolution (minimum 31 × 31 matrix size, for a field of view of 24 x 24 mm$^2$) to capture sufficient anatomical detail leading to long acquisition times. The small voxel size results in inherently low signal-to-noise ratio (SNR), making high coil sensitivity and sometimes multiple signal averages necessary, further extending experiments. These prolonged acquisition times raise the risk of motion artifacts even in anesthetized animals, often exacerbated by physiological variability, and can severely degrade data quality. Moreover, extended anesthesia may alter metabolic profiles, complicating interpretation [5]. Hardware constraints add another layer of complexity, as $^1$H-MRSI demands stronger gradients than conventional $^1$H-MRS. Finally, one additional challenge in MRSI is the need for automated and standardized processing pipelines capable of handling large datasets, performing rigorous quality assessment across thousands of spectra, ensuring reliable estimation of metabolite maps together with a precise automatic overlay of metabolic maps on anatomical images, and brain segmentation on both anatomical MR images and metabolic maps.

Approximately fifteen years ago, precursor research on magnetic resonance spectroscopic imaging (MRSI) in rodent brains utilized either the 2D-PRESS sequence (FOV = 17.6 × 17.6 mm$^2$; 1 mm slice thickness, 8 × 8 matrix size; excited volume of interest 5.5 × 5.5 × 1.0 mm$^3$; TR = 2500 ms; 21 minutes acquisition time)[6, 7] or the 2D-SPECIAL method (FOV = 24 × 24 mm$^2$; 32 × 32 matrix size; 2 mm slice thickness; excited volume of interest 10 × 2 × 10 mm$^3$; TR=1500-2500 ms; 120-135 minutes acquisition time) [8]. These techniques were notably constrained by two principal limitations: restricted spatial coverage and prolonged acquisition times. To overcome these limitations, recent studies have focused on innovative strategies such as accelerated acquisition techniques, denoising, and acquisition on ultra-high-field (UHF) scanners (>7T). For example, Wang *et al.* [9] developed the RE-CSI sequence to acquire data at 9.4T, in about 12 minutes, achieving an in-plane resolution of ~0.7 × 0.7 mm$^2$. Similarly, Simicic *et al.* [10] demonstrated the feasibility of $^1$H-FID-MRSI with substantial coverage at 14.1T in a reasonable acquisition time of ~13 minutes, and proposed the *MRS4Brain* toolbox, a pipeline for Bruker MRSI data that integrates processing, quality control, and anatomical segmentation [11]. Alves *et al.* proposed a noise mitigation method for preclinical MRSI acquisition that enables increased coverage compared to usual methods [12]. These advances highlight a promising trajectory toward faster and more robust preclinical $^1$H-MRSI; however, its application remains largely restricted to specialized studies, underscoring the need for broader accessibility and methodological refinement.

The aim of this paper is not to provide an exhaustive review of the literature on $^1$H-MRSI in preclinical research, but rather to promote broader adoption of this technique by demystifying the challenges associated with its implementation on preclinical scanners. In this guided overview, we present a practical demonstration and share recommendations based on our experience at 9.4T and 14.1T using three types of sequences (FID-MRSI, SE-MRSI, PRESS-MRSI derived from the manufacturer prototypes) to make preclinical $^1$H-MRSI more robust and accessible. We also emphasize the critical need for fast $^1$H-MRSI methods in preclinical applications to advance the research in brain metabolism in animal models.

## 2. $^1$H-MRSI data acquisition
### 2.1 Animal physiology and anesthesia

Mice and rats are the two most commonly used models for *in vivo* translational experiments due to their close genetic homology with humans, short reproductive cycle, and relatively small size. The use of animal models in high-field *in vivo* MRI/MRS enables non-invasive, longitudinal investigations of brain structure, function and metabolism with exceptional spatial and spectral resolution, thereby advancing our understanding of central nervous system processes and disease mechanisms. Deciding which model to use depends on the nature of the experiment, however rats are more commonly used in studies of central nervous system (CNS) metabolism and neurological diseases and disorders due to similarities in brain circuitry and physiology, as well as social behaviors compared to humans [5]. Furthermore, the larger brain size compared to mice makes rats a more attractive model in neuroscience.

Accurate and reproducible acquisitions in rodents depend critically on maintaining stable conditions throughout the experiments. Proper anesthesia, fixation and physiological monitoring are essential to minimize motion artifacts, preserve physiological homeostasis, and ensure reliable estimation of the metabolic profile. To ensure minimal movement and stress, appropriate anesthesia must be selected based on the animal strain and nature of the experiment. As highlighted in Lanz *et al.*, several biological and environmental factors—including stress, strain, sex, circadian cycle, body weight, and age—can modulate the physiological response to anesthesia [5]. Such variations can impact both the required anesthetic dose and the resulting MRS characterization of neurochemical profiles in the rodent brain. Because rodents can be kept for long periods of time under anesthesia, it is important to monitor physiological parameters, such as the temperature and respiration rate, during the animal setup preparation and scan. Once the animal is anesthetized, the body temperature quickly decreases. Thus, it is important to place the animal on a heating pad and continuously monitor the temperature with a probe throughout the experiment. Constant observation of respiratory frequency serves as an effective indicator for evaluating and modulating the level of anesthesia in the animal [5]. If the respiration rate becomes too low, the animal will not oxygenate sufficiently and will begin to gasp, introducing motion during the scan. Despite proper fixation of the head, breathing-related motion may produce artifacts in the cerebellum.

For non-invasive and non-surgical fixation, the head can be stabilized in a stereotaxic system using a bite bar and a pair of ear bars (Figure 1). The bite bar is first inserted into the mouth and hooked around the incisors, making sure not to catch the tongue. When securing the bite bar, it is important to make sure the head is straight and not tilted laterally along the coronal plane. Once secured, the ear bars can be positioned and slowly screwed, alternating small turns on each side to ensure the head remains straight throughout fixation. During scans, it is also necessary to cover the animal (in addition to using a heat pad) to allow it to retain its body heat.

Besides the need to fix the head of the rodent to limit movement, there are other differences and constraints to consider in preclinical imaging compared to humans. As briefly mentioned before, with proper fixation and anesthesia, rodents can be kept in the scanner much longer than humans. This is beneficial as rodents have smaller brains compared to humans; thus, in preclinical spectroscopic studies, the smaller voxel sizes needed are compensated with longer acquisition times to achieve high spectral quality and quantitative accuracy. Additionally, the anatomical difference between humans and rodents generates different RF coil filling factors for the brain, as rodents have more muscles and fat around the skull and ears that enter the field of view (FOV) of the coil, leading both to SNR limitations and possible spectral contamination. To mitigate unwanted signals, it is important that the animal is well-positioned so that the brain sits in the center of the magnet, ensuring optimal shimming in all three directions. Additionally, gently taping the ears back and taping the eyes closed with a small amount of hydration gel helps reduce possible lipid signal contaminations.

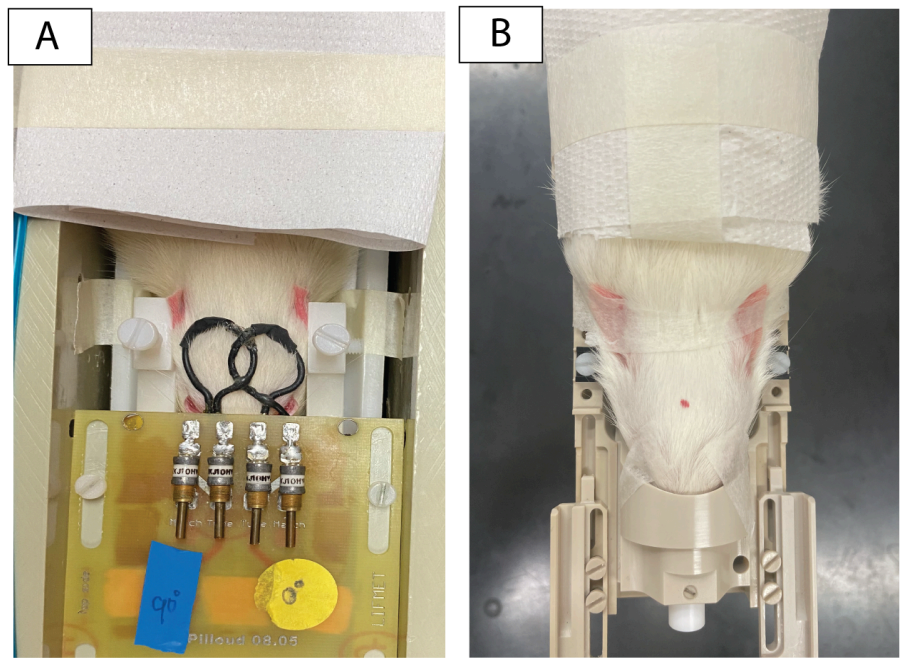

**Figure 1: Fixation of the rat using a stereotaxic system with a bite bar and ear bars. A) The setup for the 14.1T system with custom-built surface coil. B) The setup for the 9.4T system with the holder designed for the cryogenically-cooled coil.**

## 2.2 Hardware

Due to the low concentration of metabolites of interest and the small voxel volumes required to achieve sufficient high spatial resolution in the rodent brain to be able to delineate brain regions, a highly sensitive signal detection is required for small-rodent MRSI. Surface coils are a practical and effective way to achieve high SNR over a small region near the surface of the imaging subject, such as the small rodent brain [13].

In coil-dominated noise regime, typically the case for the rat and mouse brain, cryogenically-cooled coils and preamplifiers (whether transmit/receive or receive-arrays) can boost the SNR by about 2.5 to 5 fold for rodent brain MR at UHF [5, 14, 15]. However, the benefits of cryogenically-cooled coils diminishes as the static field is increasing because of increased sample thermal noise [16].

For rat brain MRSI at 14.1T, a custom-built quadrature transceiver surface coil is used (Figure 1A), comprising two oval loops (18 × 16 mm²) mounted on a curved surface of 18 × 27 mm² with a 14 mm radius of curvature. At 9.4T, our RF coil configuration consists of a cryogenically cooled 2 × 2 receive array anatomically shaped to the rat head, combined with an 86 mm inner-diameter volume transmit coil (Bruker BioSpin).

While the combination of transmit-only and receive-only coils provides a more homogeneous image, in our experience, it leads to potentially more lipid contamination compared to the surface transceiver because of the larger coverage of the receive array and the larger excitation volume of the transmit coil.

MRSI is preferably performed at ultra-high field, where the sensitivity (due to higher spin polarization) and spectral resolution (due to increased chemical shift dispersion) are improved [17–20]. However, this is counteracted by the prolonged $T_1$ and shortened $T_2$ relaxation times of the metabolites [21, 22]. Fast MRSI relies on acquisitions with a short acquisition delay (AD) or echo time (TE) to minimize $T_2$ effects and J-evolution, and thus, increase the number of detectable metabolites.

Of note, fast MRSI is generally performed using $^1$H-FID-MRSI type of acquisitions characterized by an AD (equivalent of the TE). This requires strong and fast gradients, together with high power gradient amplifiers, used here to shorten the encoding time and consequently the AD. The 9.4T setup is equipped with a B-GA12S HP gradient set (660 mT/m, 4570 T/m/s, 0.146 ms rise time), and the 14.1T with a BFG 240-120 gradient set (1000 mT/m, 5500 T/m/s, 0.243 ms rise time). Both are powered by IECO GPA-400-750 (400A/750V peak) gradient amplifiers.

## 2.3 Acquisition protocols
### 2.3.1 Choice of the MRSI acquisition sequences

The selection of the MRSI acquisition sequence must first consider the required spatial (i.e. brain coverage) and spectral (i.e. number of metabolites to be detected) information requirements, the acquisition time allocated for specific experiments in the protocol, and any challenges in acquiring that information.

Our group has implemented three general acquisition protocols based on slice-selective FID-MRSI, slice-selective SE-MRSI, and PRESS-MRSI sequences provided by the manufacturer (summary of parameters in Table 1). At UHF, FID-MRSI acquisitions are increasingly adopted as they minimize $T_2$ relaxation and eliminate J-evolution, which enhances SNR and potentially increases the number of detectable metabolites [10, 23–25]. In addition, the in-plane brain coverage is substantially increased when using FID-MRSI. Chemical shift displacement errors, which increase with $B_0$, are particularly pronounced for narrow-band, slice-selective RF pulses. Since FID-MRSI does not require refocusing pulses, in-plane chemical shift displacement artifacts are eliminated. Furthermore, its simple sequence design allows for substantial reductions in acquisition time by decreasing the TR (i.e. 811 ms at 14.1T and 813 ms at 9.4T in our protocols) while using an optimal Ernst angle excitation. In our protocols, the Ernst angle was estimated (i.e. 55° at 9.4T and 52° at 14.1T for a 0.5 ms Bruker calculated RF pulse based on the Shinnar–Le Roux algorithm with sharpness of 3, 8400 Hz of bandwidth, 50% refocusing factor) based on the shortest TR achievable with VAPOR water suppression, which typically lasts 613 ms at 14.1T or 611 ms at 9.4T, and the mean metabolite $T_1$ values at 9.4T and 14.1T, respectively [21]. The duration of the water suppression module is the main constraint for reducing TR; for example, achieving TRs of 300-400 ms would require exploring shorter water suppression schemes (e.g. WET [23–26]), which could significantly reduce acquisition time, a particularly attractive option for 3D acquisitions. One disadvantage of FID-MRSI acquisition schemes is the presence of a first-order phase in the spectra, introduced by the AD. This delay, between the excitation pulse and the start of FID acquisition, is required for spatial encoding and consists of the excitation pulse length fraction, slice-rephasing gradient and phase-encoding duration (applied concomitantly), and the dead time (ADC_INIT). Its duration depends on system performance and protocol settings, in our protocols typically ranging from 0.65 to 1.3 ms, resulting in variable first-order phase distorsions (at 14.1T for a AD of 1 ms, the first-order phase wrap is of 1080° over the 5 ppm spectral range). These phase distortions are thus dependent on both $B_0$ field and AD, resulting in spectra which exhibit dispersion-mode characteristics. Such distortions can introduce challenges for spectral fitting when using LCModel [27, 28], originally designed for single-voxel MRS spectra, for which no special handling of substantial first-order phase components is necessary. Ultra-short AD values offer significant benefits for quantifying J-coupled metabolites. They can be achieved by shortening the RF pulse and phase-encoding durations or by employing asymmetric RF pulses. In our group, the shortest AD reached at 9.4T is 0.65 ms, and is achieved by decreasing the RF pulse and phase encoding duration to 0.3 ms (bandwidth 14 kHz) and 0.2 ms, respectively (with an Auto Repetition Spoiler of 1.2 ms and 35% amplitude).

    Slice SE-MRSI (90°-180°) and PRESS-MRSI (90°-180°-180°) employ refocusing pluses, requiring longer repetition times (TR) compared to FID-MRSI (2000 ms vs ~800 ms for FID-MRSI), which resulted in substantially longer acquisition times (32 min vs 13 min). On average, the SNRt (SNRt=SNR/√(Tacq)) in FID-MRSI is about 40% higher than in PRESS-MRSI [29]. A key advantage of both SE-MRSI and PRESS-MRSI is the generation of in-phase spectra, as they are TE-based sequences, with a TE = 3.28 ms for the SE-MRSI and a minimum TE = 6-10.2 ms for PRESS-MRSI. In terms of spatial coverage, Slice SE-MRSI is comparable to FID-MRSI since the same slice is excited, whereas PRESS-MRSI is restricted to a significantly smaller rectangular volume with large chemical shift displacement errors (CSDE) at high fields. However, this strict rectangular localization in PRESS-MRSI offers the benefit of reduced lipid contamination in the acquired spectra.

    When choosing the appropriate sequence, the key factors to consider include the required metabolic information, the volume to be covered, the spatial resolution at which this information can be reliably obtained, and the acquisition time. Considering these aspects, FID-MRSI and SE-MRSI protocols provide the best brain coverage, with SE-MRSI offering in-phase spectra at the cost of a longer acquisition time.

| Preclinical MRSI Sequences | Common Parameters | 9.4T Parameters | 14.1T Parameters | Advantages | Disadvantages |
|---|---|---|---|---|---|
| FID-MRSI | **Matrix:** 31 × 31<br>**FOV:** 24 × 24 mm$^2$<br>**Spatial Resolution:** 0.77 × 0.77 mm$^2$<br>**Slice thickness:** 2 mm<br>**Acquisition BW:** 5000 Hz (9.4T) and 7142 Hz (14.1T) | **AD/TR:** 1.3/813 ms<br>**RF pulse:** 55°: 0.5ms/8400 Hz<br>**FOV Saturation:** 12 sat bands with 3-8 mm thickness<br>**WS:** VAPOR (400 Hz, last delay: 28 ms)<br>**OVS:** None<br>**total duration:** 13 min | **AD/TR:** 1.3/811 ms<br>**RF pulse:** 52°: 0.5ms/8400 Hz<br>**FOV Saturation:** 7 sat bands with 1.5-5 mm thickness<br>**WS:** VAPOR (600-660 Hz, last delay: 26 ms)<br>**OVS:** None<br>**total duration:** 13 min | + Fast<br>+ Short TE/AD<br>+ No in-plane Chemical Shift Displacement<br>+ Eliminates J-evolution<br>+ Minimizes T$_2$ signal decay | - AD induces strong first-order phase |
| PRESS-MRSI | | *n.t.<br>**TE/TR:** 10.2/2000 ms<br>**RF pulses:**<br>90°: 0.5 ms/ 8400 Hz<br>180°: 0.6 ms/ 4250 Hz<br>**FOV Saturation:** none<br>**WS:** VAPOR (400 Hz, last delay: 28 ms)<br>**OVS:** 10mm slab 1mm gap 3 ms spoiler<br>**total duration:** 32 min | **TE/TR:** 10.2/2000 ms<br>**RF pulses:**<br>90°: 0.5 ms/ 8400 Hz<br>180°: 0.5 ms/ 6800 Hz<br>**FOV Saturation:** none<br>**WS:** VAPOR (660 Hz, last delay: 26 ms)<br>**OVS:** 10mm slab 1mm gap 3 ms spoiler<br>**total duration:** 32 min | + Clean outer-volume signal cancellation<br>+ First-order phased spectra | - Longer TE → T$_2$ signal decay<br>- Longer TE → J-coupling modulation<br>- Longer acquisition<br>- In-plane CSDE (stronger in the 180° selection directions) |
| SE-MRSI | | **TE/TR:** 3.284/2000 ms<br>**RF pulses:**<br>90°: 0.5 ms/ 8400 Hz<br>180°: 0.6 ms/ 4250 Hz<br>**FOV Saturation:** 12 sat bands with 3-8 mm thickness<br>**WS:** VAPOR (BW: 400 Hz)<br>**OVS:** None<br>**total duration:** 32 min | *n.t.<br>**TE/TR:** 3.284/2000 ms<br>**RF pulses:**<br>90°: 0.5 ms/ 8400 Hz<br>180°: 0.5 ms/ 6800 Hz<br>**FOV Saturation:** 7 sat bands with 1.5-5 mm thickness<br>**WS:** VAPOR (BW: 660 Hz)<br>**OVS:** None<br>**total duration:** 32 min | + Short TE<br>+ First-order phased spectra | - Non-zero TE → T$_2$ signal decay<br>- Non-zero TE → J-coupling modulation<br>- Longer acquisition |

Table 1: Selection of proposed preclinical MRSI sequences for acquisition of brain metabolic maps, with their corresponding sequence parameters for 9.4T and 14.1T acquisitions, and respective advantages and disadvantages. Panels labelled with *n.t. (not tested) show suggested acquisition parameters. These proposed acquisition protocols were not fully tested and validated at the specific magnetic field. A more detailed table of parameters can be found in the supplemental material (Supplemental Tables 1-3)

## 2.3.2 Lipid contamination and suppression

MRSI acquisitions are highly susceptible to contamination from lipid signals originating from regions outside of the brain, such as subcutaneous fat, which can compromise the integrity of brain spectra [1, 30]. Consequently, a primary consideration in brain MRSI is the minimization of lipid signal contamination, as these signals are several orders of magnitude stronger than metabolite signals [1, 30]. In our experiments, lipid suppression is performed by saturation slabs during the acquisition and enhanced either by applying a spatial Hamming filter or by using processing-based lipid suppression tools such as $L_2$ regularization (see Section 3.2 iii). In this section, we will describe our approach to suppress the subcutaneous lipids during acquisition using saturation slabs, while in Section 4, the SVD-based processing approach implemented in the *MRS4Brain toolbox* will be briefly described.

VOI excitation by PRESS or other localisation methods, as described above, is commonly used to avoid subcutaneous lipids, although it leads to limited brain coverage in the MRSI maps. In addition, PRESS-based volume selection at higher fields leads to large CSDE. Whole-brain excitations, as well as multi-slice MRSI strongly improve the brain coverage with the disadvantage of increased subcutaneous lipid contaminations. Intermediate or long echo times mitigate the lipid contamination and baseline problems, but this limits the number of detectable metabolites (the 3 main metabolites can be detected) [1]. Short echo times or FID-MRSI type of sequences require a careful choice of the number of saturation bands and their positioning to reduce the subcutaneous fat contamination from outside the brain (Figure 2). This can be a limiting factor in terms of RF energy deposition in body tissues, in particular when using a high number of saturation bands and volume transmit coils.

For FID-MRSI and Slice SE-MRSI we are using the manufacturer implementation of "Fov Sat" available in the "Preparation" tab of any MRSI protocol. We slightly modified the manufacturer implementation by adapting the following parameters: 90° sech RF pulse, 1ms, 20 kHz bandwidth. Furthermore, in our protocols, the number of saturation slabs varies depending on the RF coil configuration and the required brain coverage. The 14.1T setup, using a quadrature transmit/receive coil, requires fewer slabs due to reduced lipid contamination from the coil (7 slabs in our protocols), whereas the 9.4T setup, with a volume transmit coil, requires more slabs to mitigate increased lipid contamination (12 slabs in our protocols) (see Sections 2.2 & 2.3.3.3 and Figure 2). Importantly, brain coverage is influenced by slab positioning; slabs placed too close to the brain, or with suboptimal transition bands, will reduce coverage. Therefore, we implemented thinner slabs with sharper transition bands, and used pairs of overlapping thin slabs in peripheral regions where extended signal cancellation was needed. This approach improved transition sharpness and minimized signal loss within brain tissue (Figure 2). The saturation slabs are carefully positioned for each experiment using the anatomical images acquired previously (Figure 2A). The efficiency of the saturation slabs can be evaluated in Setup Mode, which allows real-time modification of slab parameters. Additionally, the assessment of the fat distribution around the rat brain can be performed by acquiring a $T_2$-weighted image with two-point Dixon water-fat separation as illustrated in Figure 2B. In our experiments, the position and thickness of the saturation slabs were adjusted based on this imaging approach. Moreover, the chosen saturation band configuration can further be copied and tested on the Dixon MRI acquisition, to evaluate the effectiveness of the lipid saturation. As shown, several regions of the rat brain exhibit pronounced susceptibility to increased fat content, necessitating precise placement of the saturation slabs.

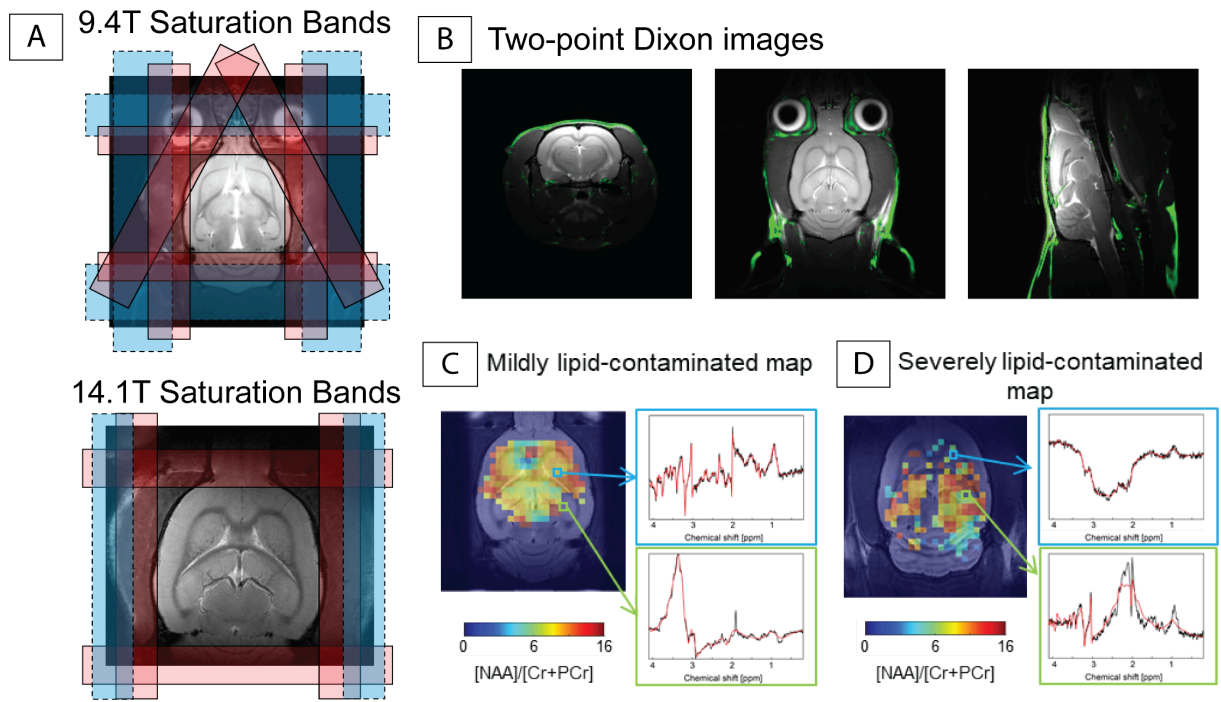

**Figure 2: A)** Saturation band placements for 9.4T and 14.1T acquisitions; **B)** Map of extracranial fat acquired via $T_2$-weighted images with two-point Dixon water-fat separation. Fat-only image in green laid over grayscale water-only image; **C)** Examples of mild lipid contamination acquired at 9.4T; **D)** Examples of severe lipid contamination acquired at 9.4T

MRSI acquisitions are usually characterized by low spatial resolutions (i.e. in the rat brain a 31x31 or 32x32 matrix size is commonly used) thus signal contamination due to the broader point spread function (PSF, more informations about the PSF can be found in the Appendix) will result in the propagation of extracranial lipid signals [1]. Due to voxel bleeding, even voxels distant from the skull can be affected [30], which can significantly deteriorate MRSI data quality (Figure 2C&D). The Hamming filter reduces the lipid contamination by reducing spatial signal spread [30], with a drawback of increased nominal voxel size from 1.21 to 1.86 [10]. Weighted *k*-space acquisitions represent a potential strategy to mitigate the effects of increased voxel size; however, this approach comes at the cost of prolonged acquisition times [1]. Similarly, higher-resolution MRSI inherently reduces lipid contamination due to a more favorable PSF [30], but also requires extended scan durations. Importantly, reconstruction errors in methods such as compressed sensing (CS) may introduce further lipid artifacts [30]. Furthermore, in our experiments the use of a spatial Hamming filter was also necessary to reduce the impact of noisy high-frequency spatial components of the signal, which are further leading to higher spectral noise and thus to spatial fluctuations in the metabolite maps when all *k*-space spatial components are equally weighted [10].

### 2.3.3 In house developed acquisition workflow

The following section outlines the acquisition protocols currently implemented in our group for MRSI data collection (videos and written protocol forms describing the acquisition protocols can be found here: LIVE Demos – MRS4BRAIN - EPFL). In general, all MRSI acquisitions follow a standardized workflow, which is summarized in Figure 3. These preparatory procedures ensure accurate spatial localization and optimal spectral quality. To reduce acquisition time per animal when needed, the MRI protocol can be shortened by omitting certain sequences or by decreasing the number of signal averages, among others. Importantly, the terminology used in this section is specific to ParaVision 360. To facilitate international harmonization, accuracy, global applicability, and straightforward implementation, all protocols described here have been developed in-house based on the manufacturer-provided protocols as the initial reference point.

As an initial step (Step 1, Figure 3 & Table 2), conventional MRI scans are performed to enable anatomical localization. When volume transmit coils are used (e.g. in our 9.4T setup), the automatic adjustments provided by the manufacturer within the Localizer sequence are applied, including RF power calibration. In contrast, when surface coils are used in transmit–receive mode (e.g., in our 14.1T setup), RF power calibration is performed manually in the Adjustment platform, as indicated in Figure 3 (dashed square). Subsequently, a $B_0$ field map is acquired in the Adjustment platform, followed by a Multislice Localizer acquisition to improve visualization and facilitate positioning of the ellipsoid for mapshim. The ellipsoid used for map shimming (depicted in green and mainly needed for the $T_2$-weighted sequences described below) was carefully positioned to maximize brain coverage while avoiding overlap with non-brain tissues. Step 1 concludes with the acquisition of anatomical 2D $T_2$-weighted images in axial and coronal orientations, which are later used for slab localization in MRSI, anatomical segmentation, and co-registration with metabolic maps.

| Sequence | 9.4T Parameters | 14.1T Parameters |
|---|---|---|
| Localizer | TE/TR = 1.9/70 ms, NA = 1, 256 × 256 matrix ||
| | Automatic ref power adjustment | manual ref power adjustment (Gauss RF pulse, BW = 6 kHz, 2 mm slice) |
| Multislice Localizer | NA = 1, 10 slices, 256x256 matrix, 24x24 FOV, TE/TR = 3/18 ms ||
| $B_0$ Map | NA=3, 96 × 96 × 96 matrix ||
| | TE1/TE2/TR = 1.75/5.3/10 ms | TE1/TE2/TR = 2.25/6.05/10 ms |
| $T_2$w Axial | NA = 2, 256 × 256 matrix, 0 gap, FOV = 24x24 ||
| | $TE_{eff}$/TR = 33/2500 ms, 28 slices, $RARE_{factor}$ = 8 | $TE_{eff}$/TR = 27/3000 ms, 20 slices, $RARE_{factor}$ = 6 |
| $T_2$w Cor | 20 slices, NA = 2, $RARE_{factor}$ = 8, 256 × 256 matrix, FOV = 24 × 24 ||
| | $TE_{eff}$/TR = 33/2500 ms | $TE_{eff}$/TR = 27/3000 ms |

| T$_2$w Atlas | TR = 4000 ms, TE$_{eff}$ = 27 ms, NA = 10, RARE$_{factor}$ = 6, 128 × 128 matrix, FOV = 24 × 24 mm, 40 slices, 0 gap, sat bands | |

Table 2: Parameters of the sequences used for adjustment, positioning and co-registration at 9.4T and 14.1T.

Step 2 contains all specific adjustments required for optimal MR spectroscopy (shimming, water suppression calibrations, positioning of the FOV saturation bands, etc), along with specific quality check procedures applied during the data acquisition (QC(A)) essential for high-quality MRSI acquisitions. The complete workflow, including these preliminary steps, is summarized in Figure 3 and detailed in the following sections.

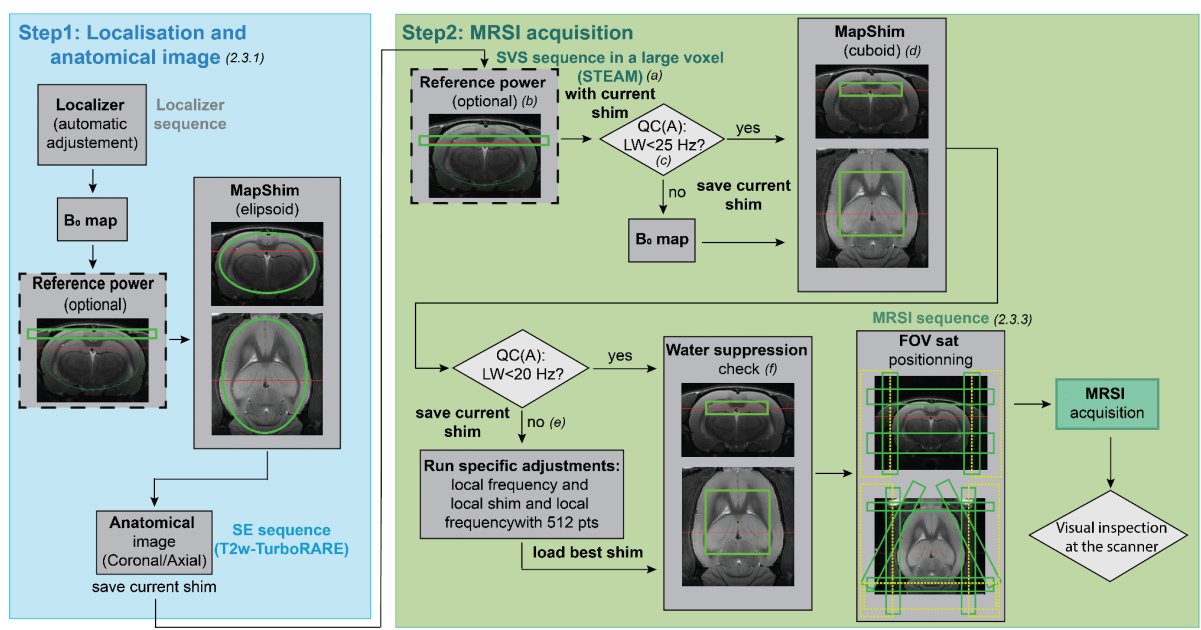

Figure 3: Standard acquisition protocol with specific adjustments for MRSI at 9.4T. Step 1 is used for acquiring the MRI data necessary for localization, anatomical reference and afterwards co-registration and visualisation of the MRSI map on the anatomical image. Step 2 includes the usual calibrations and quality checks required for MRSI. QC(A): quality check at the acquisition. Criterion for QC(A) is the linewidth of water peak to evaluate quality of the shim (B$_0$ dependent). Visual inspection consists of checking for the absence of lipid contamination in the spectrum.

#### 2.3.3.1 Acquisition for anatomical reference / coregistration

Although ¹H-MRSI acquires spectra from many spatial locations across the imaging plane, it provides limited anatomical detail. Therefore, co-registration of ¹H-MRSI data to high-resolution anatomical MRI is essential to accurately localize metabolite signals and support reliable group-level analyses. To facilitate this process, coronal and axial 2D T$_2$-weighted Turbo-RARE images are acquired for MRSI-slice positioning, shimming, and subsequent atlas-based normalization (parameters can be found in Table 2). The number and thickness of slices may be adjusted according to field strength (e.g. 40 slices of 0.3 mm thickness for 9.4T or 60 slices of 0.2 mm thickness for 14.1T). Ensuring precise overlap between the coronal $T_2$-weighted images and the MRSI acquisition plane helps minimize co-registration variability and enables accurate extraction of region-specific metabolic spectra; accordingly, the orientation of these anatomical images is used to define the MRSI slab (see Section 2.3.3.3 MRSI acquisitions and protocol description, point g) below).

### 2.3.3.2 Calibrations and shimming for MRSI

Prior to MRSI acquisitions, several calibration steps are necessary to ensure high-quality and reproducible acquired data (Step 2 in Figure 3). These steps are performed using the manufacturer STEAM sequence (adapted as shown below; more details can be found in Supplemental Table 4), chosen for its rapid acquisition and convenient display for quality assessment, as follows:

   a. After loading the STEAM sequence (number of averages (NA) =16, 1 repetition, working chemical shift 4.7 ppm, no water suppression, 2 dummy scans, TE/TR = 3/4000 ms, mixing time = 10 ms, outer volume suppression (OVS, 15/12 mm, gap = 1mm, spoiler gradient amplitude in the three directions 15-25-35% and 3 ms duration, 90°RF, 20kHz bandwidth, sech shape, 16ms duration of the module), 16 reference scans, sequence spoiler gradient amplitude in the three directions: ~25-30-20% (14.1T) and ~30-35-25% (9.4T), 90° RF of 0.5ms with same shape as for FID-MRSI mentioned above ) a VOI of 10 × 2 × 10 mm$^3$ is positioned in the rat brain at the target slice within the region of interest - based on our experience this voxel size covers the essential part of the MRSI slice sensitive volume covered by the surface coils, as determined by the quality of the water linewidth and metabolite SNR. In Figure 3, this VOI includes primarily the hippocampus in the axial orientation (2 mm) and the hippocampus, cortex, and striatum in the coronal orientation (10 × 10 mm$^2$). This VOI can be positioned in other brain regions according to the requirements of the study. Additionally, the NA in the STEAM sequence may be reduced to two to shorten acquisition time when necessary.
   b. As mentioned above, for transmit–receive surface coils only (e.g., 14.1T), RF power calibration needs to be performed in the Adjustment platform by positioning a slab within the slice we plan to measure (Figure 3, dashed square).
   c. Optimal $B_0$ shimming over the MRSI slice can be achieved in a sequential way [31]. The spectroscopic signal quality, after the ellipsoid shimming, can be quickly checked with a water signal acquisition over the 10 × 2 × 10 mm$^3$ VOI, representing the central part of the MRSI matrix. If the water linewidth exceeds 25 Hz at 9.4T, a Localizer sequence is loaded and a second $B_0$ map is acquired in the Adjustment platform. If the linewidth is below 25 Hz at 9.4T, no additional $B_0$ map is needed, and the workflow proceeds directly to point d). It is noteworthy that this linewidth is an approximation of achievable MRSI signal linewidth, since it can be seen as a $B_1$-weighted sum of the individual MRSI voxel signals. Furthermore, the total STEAM voxel linewidth is also affected by the individual MRSI voxels' frequency shifts linked to $B_0$ inhomogeneities (Figure 4). Alternative shim geometries or sizes may be feasible and they represent a promising avenue for future studies ; however, these configurations were not evaluated. They should provide both adequate spatial coverage and temporal stability.
d. For improving the shim quality a 2$^{nd}$ STEAM sequence is used with a MapShim volume centered on the 10 × 2 × 10 mm$^3$ voxel with iterative corrections. A water linewidth below 20 Hz is targeted at 9.4T for the voxel size of 10 × 2 × 10 mm$^3$ (below 30 Hz for 14.1T).
e. If this target value is not reached then a third STEAM sequence is loaded and some specific adjustments are performed in the Adjustment platform (Local frequency, Local Shim and Local frequency using only 512 points in the FID). Shimming parameters convergence is sometimes hard to achieve over large volumes, especially when reaching tissue interfaces such as in the cortex. Better linewidth results, measured in the 10 × 2 × 10 mm$^3$ STEAM volume, can be achieved when shimming over a slightly reduced volume, such as 9 × 2 × 9 mm$^3$. Furthermore, some brain areas are easier to shim than others.
f. Once the target water linewidth is achieved, the next step is to test and calibrate water suppression (WS) using the VAPOR module on the same VOI. Similar to shimming, this calibration provides an approximation of the achievable water suppression for MRSI. For that, the STEAM sequence from point e) is duplicated and VAPOR model is enabled. In our protocols, we are using the following parameters in the VAPOR module, which are slightly changed from Bruker implementation: Hermite RF pulses, bandwidth: 350 Hz (14.1T) and 270-300 Hz (9.4T), flip angles 1 and 2: 84°/150°, last delay 22 or 26 ms (14.1T) and 26 or 28 ms (9.4T), 634 ms duration (14.1T) and 637 ms duration (9.4T)). This choice of parameters allows a robust water suppression with only two parameters to calibrate: the water suppression bandwidth and the last delay. These parameters are calibrated by running the sequence in a Setup Mode. If needed, a short acquisition of metabolite signal is performed, in our specific case 32 averages are used (they can also be done as repetitions, and thus, the signal is stored by individual shots for future preprocessing steps: B$_0$ drift corrections, etc [32]). A complete description of parameters can be found in Supplemental Table 4. A water signal is also acquired for quantification purposes as described at point k) below (same parameters as for the metabolite acquisition except that 16 averages are used).

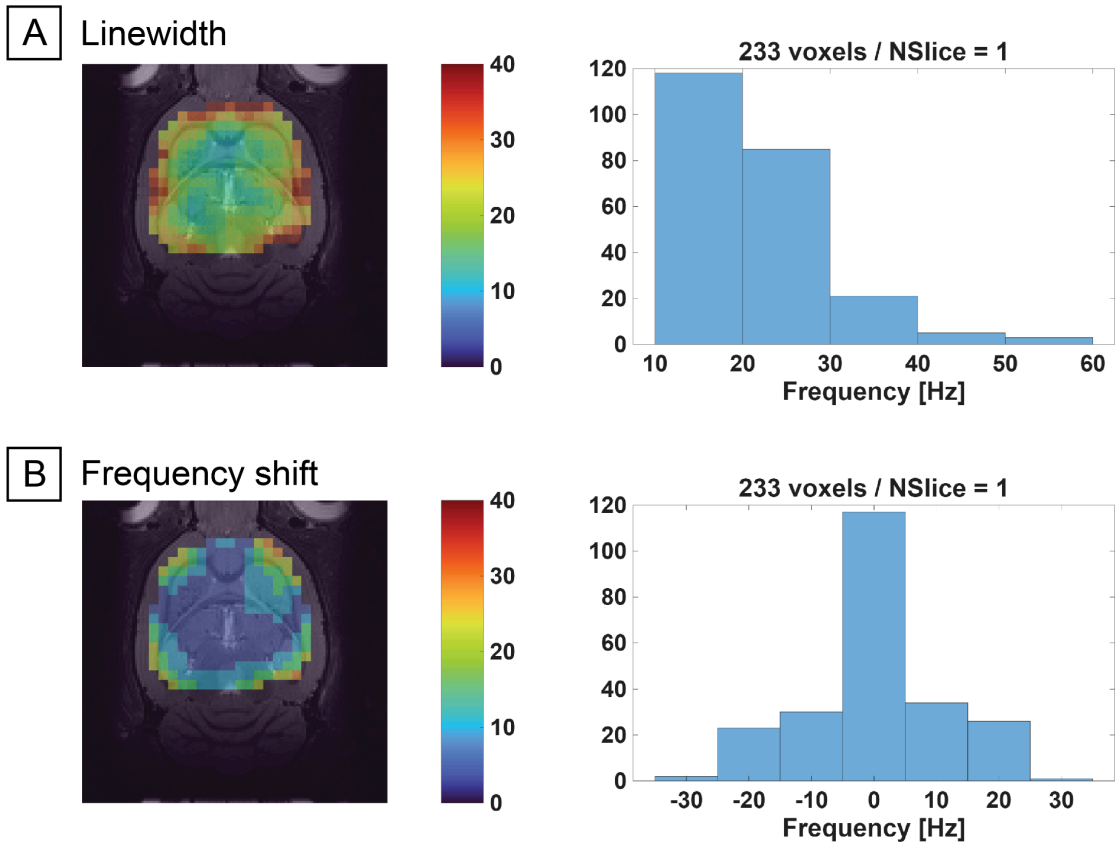

**Figure 4: Water peak linewidth (A) and water peak frequency shift (B) at 9.4T visualized as a map (provided by the *MRS4Brain Toolbox as quality control checks*) and as an histogram representing the frequency over 233 voxels on one slice. The frequency shift is represented on the histogram, while the absolute value of the shift is shown on the map.**

**2.3.3.3 MRSI acquisitions and protocol description**

Finally, the in-house developed MRSI protocol (adapted from Bruker sequences, and using the sequences described above in Section 2.3.1) is loaded (LIVE Demos – MRS4BRAIN - EPFL) and the following steps described below are performed. Details on the type of the sequence and specific parameters can be found in the MRSinMRS [33] (Supplemental Tables 1-3) and in the Section 2.3.1 above.

    g. Copy the slice orientation (not geometry) from the coronal $T_2$ weighted acquisition to ensure a direct coregistration between the anatomic and metabolic acquisitions

    h. Set the same slice offset (coordinate along the perpendicular axis to the plane) as in the previous STEAM acquisition.

    i. Adjust the position of the saturation slabs using the coronal and axial $T_2$ weighted images acquired previously. Adapt the number of saturation slabs and their position based on the RF coils used (see above Section 2.3.2 and Figure 2). At 14.1T we are currently using 7 saturation slabs to minimize lipid contamination (thickness of the slabs between 1.5 to 4 mm, slabs 2&3 and 4&5 saturate the same region, 12-13 ms duration, Auto Spoiler [0.86 ms

duration and 13% amplitude] [10]. At 9.4T, due to the usage of a volume coil for excitation, we are using 12 saturation slabs with the same parameters as at 14.1T with the following exceptions: their thickness is ranging from 3 - 8 mm and the Auto Spoiler [0.8 ms duration and 40% amplitude] leading to a duration of the module of 22 ms [10]. Importantly the number and thickness of saturation slabs determine the duration of the "Fov Sat" module, which is currently implemented by the manufacturer between the final water suppression RF pulse and the start of the MRSI sequence. A too long module will impact the water suppression efficiency. In our protocols at 9.4T a maximum interval of 28ms (PVM_VpInterPulseDelay number 7) between the last water suppression RF pulse and the onset of the MRSI sequence is considered acceptable to maintain adequate water suppression.

j. Adjust the VAPOR water suppression using the calibrated values obtained under point (f) above (e.g., bandwidth and final delay; of note the water suppression bandwidth is larger for MRSI than for STEAM sequence: 600-660 HZ at 14.1T (module duration 613 ms) and 400 Hz at 9.4T (module duration 611 ms)). Water suppression can be verified and further refined in Setup Mode. Note that the Setup Mode display reflects suppression quality across the entire slice without phase-encoding gradients; therefore, it does not fully represent the quality achievable in individual MRSI voxels.

k. Acquire the metabolite and water signal separately (working chemical shift 2.7 ppm); for the water signal two options can be used: i) fully disable the VAPOR module; or ii) deactivate only the RF pulses within the VAPOR module while keeping the module enabled, ensuring that the gradient components remain active.

### 2.3.4 *In vivo* macromolecules acquisition

Short TE $^1$H MR spectra contain contributions from the broader signals of macromolecules (MM). Following experts' consensus [19], for optimal metabolite estimation, the MM contribution should be included in the basis set used for quantification. The double inversion recovery provides good MM signal recovery with excellent metabolite suppression. Because the rodent brain is mainly composed of grey matter and doesn't appear to differ across brain regions, a single MM spectrum was acquired for the basis set [19, 34]. A double inversion recovery module should be programmed into the sequence in use (i.e. for $^1$H-FID-MRSI at 14.1T: **TI = 2200/850 ms**, 15 × 15 matrix size, 6 averages, TR = 3400 ms, HS1_R20.inv RF pulse of 2 ms duration (180°), 2048 FID data points, 7 kHz acquisition bandwidth, RF excitation was performed with a 90° pulse due to increased TR, total acquisition time of ~76 minutes; at 9.4T the only difference was the 5kHZ acquisition bandwidth) [10]. Due to lower SNR of the acquired MM, a smaller matrix size was used while at least 2-7 voxels should be summed to obtain the final MM signal (2-4 voxels were summed at 9.4T, while 6-7 voxels were summed at 14.1T [10]), and the residual water and metabolite signals (tCr [~3.88 ppm], Glu + Gln [~3.72 ppm], Ins [~3.50 ppm], Tau [~3.39 ppm], Tau + tCho [~3.21 ppm], NAA [~2.64 ppm], Glu [~2.32 ppm], Gln [~2.09 ppm]) should be removed using the AMARES [35] module in jMRUI [36]. The final MM signal is to be included in the basis set. An example on how to process MM spectra can be found online on our group webpage: LIVE Demos – MRS4BRAIN - EPFL

## 3. Processing workflow

In our group, the entire processing-fitting-quantification pipeline is automated using the *MRS4Brain* toolbox [11] ([MRS4Brain Toolbox – MRS4BRAIN - EPFL](#)). The main steps used in our MRSI pipeline are illustrated in Figure 5 and briefly described below.

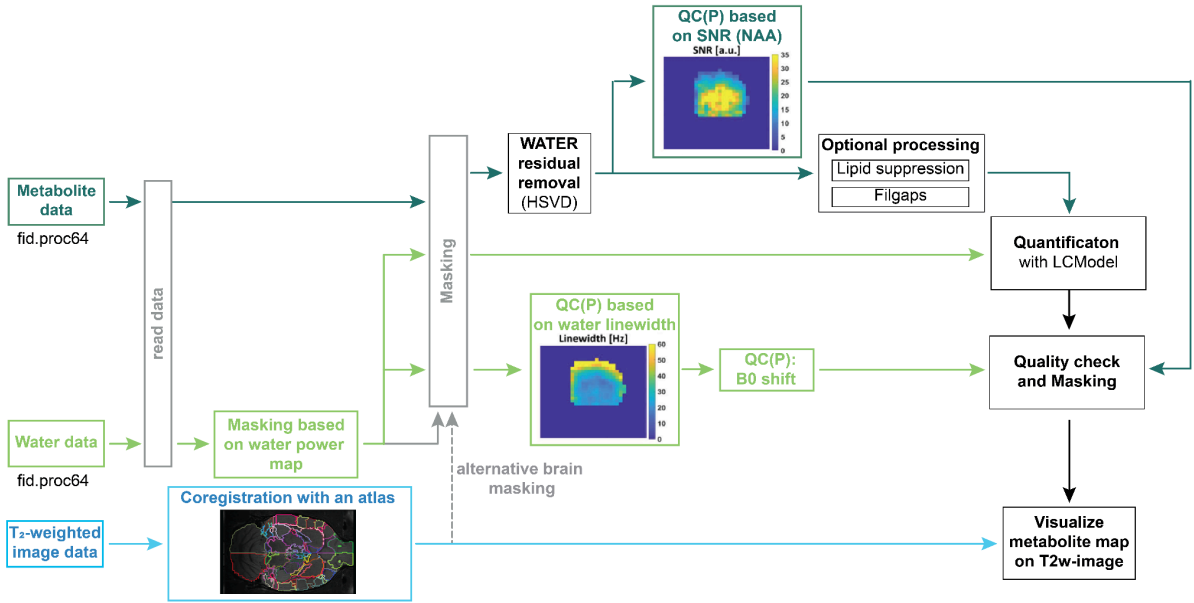

Figure 5: Overview of our MRSI processing pipeline used in *MRS4Brain* toolbox, illustrating standard data processing steps and quality checks performed (QC(P) - available in the "Show Results" tab of the toolbox). Some quality check procedures can also be used after LCModel fitting (SNR(NAA), CRLBs, LCModel linewidth - available in the "Display Settings" tab of the toolbox) to generate masks for visualization purposes and analysis.

### 3.1 Coregistration with an atlas

Brain registration with an appropriate anatomical segmentation is paramount for region specific MRSI analysis, available in the *MRS4Brain* toolbox. Because image localization and quality vary with research objectives, animal strain, and scanner infrastructure (e.g. 9.4T or 14.1T), a field-strength-specific anatomical template is strongly recommended. This template can be generated from $T_2$-weighted images acquired across a cohort of animals used in the study. Then, the specific brain atlas can be aligned to this custom template to enable brain segmentation

To avoid any alteration of spectral signals due to registration procedures, it is preferable to align the template to the individual animal's image rather than the inverse. In this workflow, the homemade template is nonlinearly coregistered to the coronal anatomical image using a dedicated processing tool such as ANTs [37], employing both affine and symmetric image normalization transformations. The resulting transformation matrix is subsequently applied to transfer the atlas brain labels from the template space into each animal's anatomical space, where they are used to define the spectral regions of interest. Currently, the *MRS4Brain* toolbox [11] provides two custom rat brain templates for the 9.4T and 14.1T field strengths. These templates were generated from $T_2$-weighted Turbo RARE images and include segmentations derived from the SIGMA rat brain atlas [38].

## 3.2 Processing MRSI data
### i) Reading of MRSI data file using the MRS4Brain toolbox

Data reconstructed in the image space can be loaded via the fid_proc.64 file (or via fid file in versions of PV360 prior to v.3.X). An in-plane two-dimensional *k*-space Hamming filter, implemented in PV360 was consistently applied along each spatial dimension during reconstruction. Due to the structure of these files ($N_{FID}$ X $N_{Enc}$ where $N_{FID}$ is the number of spectral points and $N_{Enc}$ is the number of encoding steps acquired), the dataset is restructured to match the 3-dimensional object structure desired for 2D-MRSI processing procedures (4-dimensional object for 3D-MRSI). The Bruker provided reconstruction for MRSI acquisitions accounts for the averaging of the data if acquired with more than one average and for the possible Fourier transform shifts. The first 77 points of each FIDs are cut as they represent the delay of the digital filtering within the hardware (the group delay points), but to preserve the spectral dimension zero-filling is applied with the same number of points cut. The FIDs are then corrected with respect to the frequency shift (working chemical shift on the scanner) applied during the acquisition. These steps are reproduced for both water and metabolite signals. The water MRSI dataset is also scaled to the same receiver gain value used for the metabolites.

### ii) Quality Checks and Masking in Image Space

The first step is to select the brain region using either the coregistration presented in Section 3.1 or a water-power map, typically by applying a threshold of about 0.5 times the mean water signal. Once the brain is extracted, the quality of the data is assessed through three checks, collectively referred to as QC(P): linewidth of the water peak, SNR of metabolites (more specifically the NAA), and shift of the water peak (used only as an additional verification of data quality) (see Figure 5).

Ensuring good shim quality is essential for accurate quantification. A common criterion for acceptable shimming is 0.05 ppm (below 20 Hz for 9.4T and 30 Hz for 14.1T) [5]. Based on this threshold, a water linewidth mask can be applied to retain only data with sufficient shim quality (not implemented in the toolbox). Additionally, the global linewidth provided by LCModel is used as masking in the toolbox. Finally, an additional mask is applied based on the SNR of the NAA peak (or LCModel SNR estimate in the toolbox), which should typically be greater than 10 (4 for LCModel SNR).

### iii) Optional processing step: retrospective lipid suppression ($L_2$ regularization)

Although during data acquisition, saturation slabs were applied to suppress cranial lipids, lipid contamination may potentially occur in spectra, particularly for voxels located near the skull. To mitigate this, we applied retrospective lipid suppression based on singular value decomposition (SVD), which assumes that lipid and metabolite signals are orthogonal in the time or frequency domain and exhibit no spatial overlap [39]. Brain and scalp regions were first delineated using a water power mask. SVD was then performed on the scalp voxels to derive an orthogonal basis representing the lipid components. The rank of this basis was selected by evaluating the energy ratio between brain and scalp regions following application of the corresponding projection operator: EBrain/ESkull ≥ α (α = 0.8 for our data). Lipid removal using the $L_2$ regularization method can induce errors in NAA estimation due to closer vicinity of NAA signal to lipids and baseline distortions occurring at 2 ppm below the NAA singlet peak [40]. Furthermore, the combination between saturation slabs, *k*-space filtering, and the SVD-based lipid suppression must be carefully evaluated, as their interaction may introduce distortions in the MM pattern. Such distortions can lead to inaccurate MM estimation during the fitting procedure, ultimately resulting in errors in metabolite estimates.

*iv) Ongoing processing implementations*

Beyond the Bruker provided reconstruction, the rawdata.job0 file provided after each acquisition gives the user freedom in how to reconstruct the MRSI dataset to fit its own desire and needs. Our job0 reconstruction takes the raw signals acquired from individual coils and combines them to generate a single complex signal for each *k*-space point. This combination follows the principles implemented in the standard Bruker reconstruction pipeline, including the application of predefined scaling factors and phase corrections stored in the method file. The purpose of replicating the Bruker pipeline in our code is to provide full transparency and flexibility, enabling verification of each processing step and assessment of data quality throughout the workflow. Alternatively, coil combination can be performed using an SNR-based weighting approach, as implemented in FID-A [41] (*op_combineRcvrs*), which optimizes the contribution of each coil based on the relative phases and amplitudes determined in the time domain on unsuppressed water data [42]. For cases where multiple averages are acquired for the same *k*-space point, phase and frequency alignment can be performed prior to summation, in accordance with Near et al. [32].

**3.3 LCModel fitting and quantification**

The spectra contained in the brain region were quantified using LCModel incorporated in the *MRS4Brain* toolbox (Figure 6). Sequence-specific basis sets of metabolites can be simulated using NMRScopeB [43] from jMRUI using published values for J-coupling constants and chemical shifts [44, 45] with sequence parameters corresponding to the *in vivo* metabolite acquisitions (some examples are provided: MRS4Brain Toolbox – MRS4BRAIN - EPFL). In each basis set, the following metabolites should be included: aspartate (Asp), ascorbate (Asc), creatine (Cr), phosphocreatine (PCr), γ-aminobutyrate (GABA), glutamine (Gln), glutamate (Glu), glycerophosphorylcholine (GPC), glutathione (GSH), glucose (Glc), inositol (Ins), N-acetylaspartate (NAA), N-acetylaspartylglutamate (NAAG), phosphorylcholine (PCho), phosphorylethanolamine (PE), lactate (Lac), and taurine (Tau). PCho and GPC, and Cr and PCr were expressed as tCho (PCho + GPC) and tCr (Cr + PCr) due to better accuracy in the estimation of their concentration as a sum. The MM spectrum, acquired as described above, should also be included in the basis set.

Typically, spectra are fitted on a frequency range between 0.2 and 4.2 ppm, as most common metabolites of the brain resonate within this range. However, when using FID-MRSI acquisitions, lipid contamination can pose challenges, as lipids resonate between 0.9-1.5 ppm, overlapping with MM signals and other metabolites of smaller concentrations. This is common for voxels located in the periphery and may lead to an overestimation of MM, thus affecting the quantification of the other metabolites. Furthermore, the residual signal will be increased in LCModel fitting leading to a lower SNR estimation, as LCModel calculates SNR by taking the ratio of the amplitude of the NAA signal and the residual signal [12, 28]. With low SNR, there will be a loss of voxels in the metabolic maps that don't pass the QC(P). One potential approach to mitigate this issue is to restrict spectral fitting to a narrower frequency range (1.8-4.2 ppm) and employ a basis set that includes a fitted MM signal [46]. However, this strategy might underestimate MM contributions, as the prominent MM signal near 0.9 ppm is excluded, and metabolites such as lactate and alanine cannot be quantified. On the other hand, limiting the fitting range reduces residuals, thereby improving LCModel estimation of SNR and lowering CRLB estimation.

Relative concentrations can be reported with respect to tCr; however, caution is needed as tCr is often altered in disease models [47]. Reporting absolute concentrations from FID-MRSI is more challenging due to the first-order phase introduced by the AD. One approach to address this issue is applying Back-Linear Prediction of missing FID points (see Section 4.1) [27].

The *MRS4Brain* toolbox provides multiple simulated metabolite basis sets that incorporate experimentally measured MM contributions [11] (MRS4Brain Toolbox – MRS4BRAIN - EPFL). These basis sets are generated specifically for each acquisition protocol using the corresponding sequence parameters. Furthermore, for our data we are using the following fitting strategies: a) FID-MRSI: narrower frequency range between 1.8–4.2 ppm; b) SE-MRSI, PRESS-MRSI, Single-voxel spectroscopy (SVS): fitted on a frequency range between 0.2 and 4.2 ppm (Figure 6).

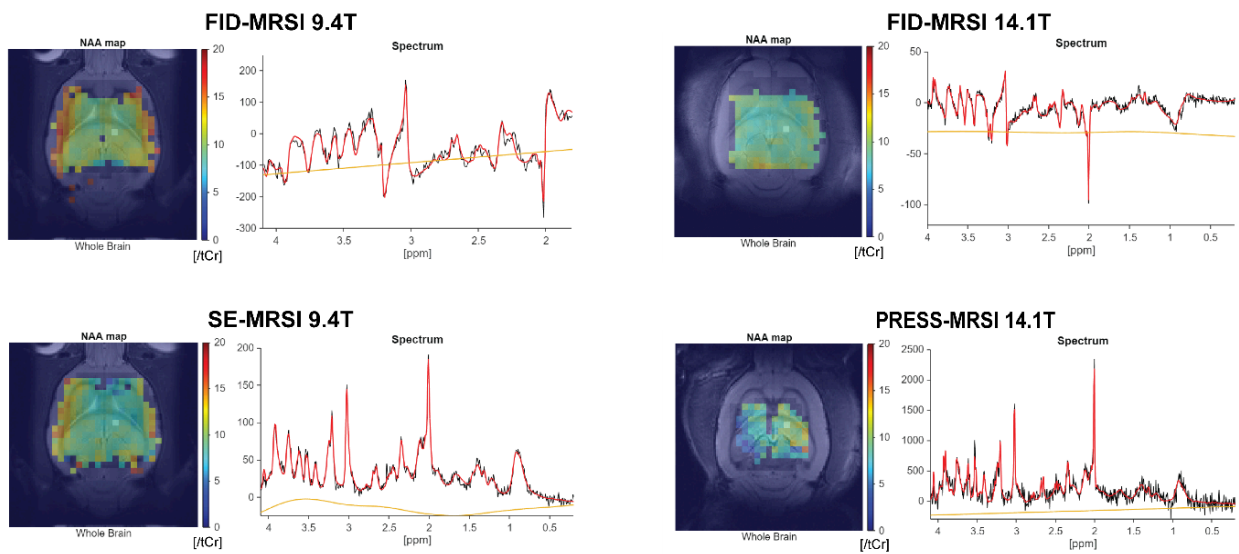

**Figure 6: Representative metabolic maps of NAA acquired using FID-MRSI at 9.4T, PRESS-MRSI at 14.1T, and SE-MRSI at 9.4T, and a respective spectrum from a voxel in the hippocampus**

## 4. Advanced tools and applications for MRSI
### 4.1 Back-prediction of the first missing FID time-domain points

A $^1$H-FID-MRSI processing strategy to address the challenge imposed by the AD (described above) is the Back-Linear-Prediction (BLP) methodology, consisting of the back-prediction of the first missing FID time-domain points up to AD = 0 ms via autoregressive reconstruction methods. Indeed, due to the presence of an AD for the phase-encoding stage, the first FID points are not acquired, causing loss of high-amplitude initial signal components. Besides allowing for such a recovery, the BLP to AD = 0 ms allows for cancelling spectral first order dephasing, mitigating LCModel quantification biases induced by different first order phase conditions related to specific AD values [27]. In this way, such a methodology introduces a common quantification framework (i.e., AD = 0 ms) convenient for comparisons between studies acquisitions at different ADs, representing an important step towards MRSI standardization. Furthermore, this approach can facilitate the interpretation of MRSI spectra, allowing for an easier identification of spectral peaks, amplitude variations, and potential contaminations or distortions. The MRSI spectra are restored to the same phasing state as for SVS quantifications, where standard quantification methods have been validated for their reliability and performance.

A recommended application setting consists of applying an autoregressive algorithm (e.g., *fillgaps* or *arburg* MATLAB functions, based on Burg's method [48]) to each FID signal up to AD = 0 ms, in the real-space MRSI matrix (already processed). The corresponding spectra can then be quantified in LCModel, using a basis set with components simulated at null AD and a SVS-acquired MM signal at very short TE (TE = 3 ms) [27], as proxy for a phased MM profile virtually acquired at AD = 0 ms. Future inclusion of BLP reconstructed FID-MRSI MM signals will require further validation. Also, a sufficiently flexible baseline fitting (LCModel parameter *dkntmn* = 0.15) is recommended, to account for possible baseline distortions that may arise by the rephasing of residual contamination contributions in the BLP process.

The BLP procedure is implemented as an optional processing step in the MRS4Brain Toolbox, where it can be activated in the *Processing Steps* panel, ensuring straightforward usage [11].

### 4.2 Acceleration via Compress Sensing

Traditional phase-encoded $^1$H-MRSI is limited by its long acquisition time. Standard application following the parameters suggested in Table 1 lead to a minimum of 13 minutes per acquisition for FID-MRSI. One possible strategy to mitigate this in preclinical applications is the CS acceleration scheme, which works by undersampling the *k*-space during the acquisition allowing an acceleration inversely proportional to the percentage of acquired *k*-space measurements. CS stands out due to its ease of application, its ability to be combined with different reconstruction tools and its exemption of *g*-factor penalty [4, 24, 39]. However, this acceleration technique comes with a price with regards to spatial resolution, as changing the sampling pattern causes some issues with noise-like aliasing and increase in lipid suppression [24]. Due to these drawbacks, CS has to be used with careful planning when applied to MRSI acquisition.

The Bruker implementation of CS allows the user to interact with the core fully sampled at the center of the *k*-space, the undersampling factor (AF-acceleration factor) and with a set of predefined reconstruction parameters based on Lustig *et al.* method [49]. The reconstructed fid_proc.64 files stored after acquisition can be directly uploaded in the *MRS4Brain* toolbox. The effects of all these parameters on the PSF have to be taken into account [50]. In our group we have recently tested the CS technique and our results at 14.1T and 9.4T demonstrated that it is feasible to reduce the acquisition time from 13 min to 3.25 min using an AF = 4 and core size of 20%. CS-FID-MRSI yielded results comparable to standard FID-MRSI across acceleration factors, with minor precision loss. For the Core parameter, values below 10% led to reduced map coverage due to increased lipid contamination and metabolite map granularity. The application of CS is mostly recommended for 3D acquisition where the high number of phase-encoding steps renders MRSI acquisition even longer, however CS remains limited in terms of acceleration factors (4 times faster before contamination) [50, 51].

### 4.3 Increased brain coverage via 3D MRSI

Extensions of the MRSI technique can be done in order to acquire 3-dimensional metabolic maps, an option available on PV360 v3.5 and implemented in our experiments (FOV of 24 × 24 × 9 mm$^3$ for a matrix resolution of 31 × 31 × 9, implying 9 slices of 1 mm thickness in the slice encoding direction) [50]. This can extend quite substantially the coverage of the MRSI technique while also allowing for an SNR increase (on average, almost twice the value compared to 2D-MRSI due to a larger excitation volume 9 mm vs 2 mm [50]). However, 3D-MRSI requires much longer acquisition time due to the 3rd phase encoding gradient required in 3D settings (additional 9 phase encoding steps in slice encoding direction) and it is generally advised to combine acquisition with an acceleration scheme such as CS or spatial-spectral encoding. In our implementation we used CS with AF=4 and core size=3% - adapted automatically by the manufacturer while allowing for an effective *k*-space pattern more uniformly distributed on the volume of interest, with a reduction of the total acquisition time from 118 min to approximately 29 min [50].

The 3D-MRSI protocol requires adaptation in the shimming procedure (described in Section 2.3.2.2.) in order to achieve an acceptable spectral linewidth throughout the whole volume: the VOI used for the MAPSHIM step of the procedure is slightly smaller than the 3D-MRSI FOV in the slice encoding direction (using the suggestion from Section 2.3.2.2., the VOI would go from 10 × 2 × 10 mm$^3$ to 10 × 8 × 10 mm$^3$ where 8 is a thickness smaller than what is acquired with the 3D-MRSI in the slice encoding direction). As expected, increasing the VOI resulted in a slight broadening of the water linewidth measured with the STEAM sequence, as MAPSHIM encountered greater difficulty achieving optimal shimming for larger volumes. Nevertheless, this 3D methodology provides a noteworthy improvement in SNR, while delivering metabolite estimates comparable to those obtained with 2D FID-MRSI and offering superior spatial coverage.

Importantly, the saturation bands need to be adapted as there is a higher risk of lipid contamination due to the larger excitation volume. An additional band is recommended on the top of the cranium of the rat in order to limit through-plane contamination of the slices [50]. Furthermore, the additional phase encoding in the slice direction renders the 3D modality more sensitive to PSF related issues due to a lower number of steps and thus through-plane voxel contamination.

**4.4 Increased brain coverage Multislice MRSI**

An alternative for 3D metabolic maps is Multislice MRSI, a protocol implemented in our group with the parameters described below [50]. Instead of using phase-encoding to acquire the third spatial dimension, Multislice MRSI sequentially excites a user-defined number of slices to extend the coverage along the slice selection direction. This approach increases the TR proportionally to the number of slices ($TR_{Multislice}$ = 9 × 822 ms; 822 ms = $TR_{FID-MRSI}$), requiring adjustments to the flip angle (90° instead of Ernst angle) and resulting in longer acquisition times compared to 2D MRSI (29 minutes with CS for multislice FID-MRSI *vs* 13 minutes for 2D FID-MRSI without CS). CS was thus applied per slice (AF = 4 and a Core = 20% with an effective *k*-space center having the shape of a parallelepiped whose length is on the slice selection direction, thus not as uniformly distributed as for 3D-MRSI). The same configuration for the saturation bands is used as for 3D-FID-MRSI. Finally the nominal voxel size, compared to 2D, was reduced from 1.19 µL to 0.59 µL for both 3D and Multislice FID-MRSI.

Additionally, this difference in encoding process has consequences on the spatial resolution: as the slice selective pulse determines the slice thickness, the PSF issues in the through-plane direction are less pronounced than for phase-encoded 3D-MRSI (see Appendix PSF).

**4.5 Indirect detection of deuterium**

Deuterium metabolic imaging (DMI) enables the mapping of key metabolic pathways in the brain by tracking the metabolism of deuterium-labelled tracers, such as glucose [52–56]. As an interesting alternative, indirect detection of $^2$H using $^1$H-MRSI to quantify $^2$H turnover through the loss of proton signal when $^1$H is exchanged with $^2$H. This method, known as Quantitative Exchange Label Turnover (QELT) [57, 58], does not only provide the possibility to perform $^2$H labelling studies without the need for specific X-nuclei hardware, it can also make use of the latest $^1$H-MRSI approaches to quantify both labelled metabolites and their total pool size in a single measurement, as well as the full $^1$H metabolic profile, using the same internal reference [59].

The recent preclinical extension of $^1$H-FID-MRSI, described above, enabled a characterization of regional glucose oxidative metabolism, under continuous and controlled infusion of [6,6'-$^2$H$_2$]glucose, resulting in a quantitative description of local TCA cycle flux with minimal modelling assumption [60]. This method can virtually be adapted to all $^1$H-MRSI protocols described above (2D, 3D), under the conditions that a sufficiently high temporal resolution can be achieved to follow and characterize the dynamics of the labelled downstream metabolites, and that enough SNR is provided for robust measurements of $^1$H signal losses. For the case of deuterated Glc and the progressive labelling of Glu and Gln (or their sum Glx), a typical temporal resolution of 10-15 minutes is required.

## 5. Future Developments

*Increase coverage*

Achieving adequate in-plane coverage remains a significant challenge for MRSI on preclinical scanners. Coverage limitations can arise from multiple factors, including suboptimal shimming at the skull edges, and the use of saturation bands to suppress lipid contamination. Despite rigorous efforts to optimize $B_0$ shimming, correcting $B_0$ inhomogeneities across an entire brain slab remains a critical issue. Future research should focus on developing advanced shimming strategies [31], such as real time $B_0$ correction [61], to improve field homogeneity across large brain slabs or large shim volume particularly at peripheral regions. A second major constraint is the reliance on multiple saturation bands to mitigate lipid contamination originating from extracerebral regions. One strategy to reduce lipid contamination involves modifying the PSF by applying a Hamming filter with weighted averaging during acquisition; however, this approach substantially increases acquisition time as a trade-off. Alternative lipid suppression techniques that reduce reliance on multiple saturation bands, such as selective RF pulse design, should be explored [62]. Hardware innovations also represent a promising avenue; for instance, designing cryoprobes with integrated transmit-and-receive surface coils for rats could replicate the improved coverage observed with mouse cryoprobes, although such configurations are not currently available. Local shimming coils, placed close to the rodent head, can generate highly localized magnetic field corrections that compensate for the strong and steep susceptibility variations found in the rodent brain that cannot be adequately corrected by the scanner's spherical harmonics shim system [63]. Future studies should also investigate the efficiency of lipid suppression using metabolite-lipid orthogonality and the best compromise between the amount of lipid contamination (depending on the RF coils used and presence or not of saturation slabs in the sequence) and the rank of the basis for the approximated lipid subspace with and without saturation slabs.In this paper, we mainly focused on 2D $^1$H-MRSI; however, a substantial improvement in spatial coverage with a good SNR could be achieved through the implementation of 3D $^1$H-MRSI. This technique is widely adopted in clinical practice using spatial spectral encoding, such as EPSI and concentric rings (CRT) methods, and offers significant potential for advancing investigations of region-specific metabolic processes [4, 64]. Nonetheless, its implementation poses considerable challenges, as the acquisition time, proportional to the repetition time and the number of phase encoding steps, increases markedly with this approach. Future work should therefore prioritize the integration of advanced acceleration strategies, new shimming procedures and robust reconstruction algorithms to enable feasible 3D MRSI acquisitions within a reasonable duration (e.g., <20 minutes) without compromising spectral quality.

*Acceleration method*

In the context of preclinical MRSI development, acquisition speed constitutes a major limitation. Accelerated data acquisition is essential for mitigating physiological artifacts, which is particularly critical for applications such as monitoring dynamic metabolic processes or conducting extended protocols, including diffusion-weighted MRSI. Strategies for improving acquisition efficiency encompass reducing repetition times via shorter WS modules and optimizing *k*-space sampling schemes (see Bogner's review on acceleration [4]), as well as advancing the acquisition methodologies outlined in Section 4. Non-linear sampling techniques employing non-Cartesian trajectories, such as CRT, spiral, and radial schemes, offer substantial acceleration while preserving spectral fidelity. Furthermore, the integration of CS with CRT represents a promising approach to enhance acquisition speed without compromising spatial or spectral resolution [65].

For sequences other than FID-MRSI, the incorporation of selective excitation pulses may further decrease acquisition time by obviating the need for water suppression, thereby enabling shorter TR [62]. This strategy also contributes to the reduction of lipid contamination.

*Denoising*

Given the inherently low SNR in MRSI acquisition data, denoising represents a critical step toward improving data quality. Preliminary investigations in our laboratory have demonstrated that the Marchenko–Pastur principal component analysis (MP-PCA) denoising method is a promising approach for achieving more accurate metabolite quantification, with reductions in standard deviation of several percent observed in the hippocampus [12]. This approach appears to outperform low-rank total generalized variation reconstruction techniques. For other nuclei and acquisition contexts, such as dynamic $^2$H-MRSI, additional advanced denoising strategies have been proposed, including SPIN-SVD and tMPPCA [66]. These methods warrant further evaluation in preclinical imaging settings.

*Toward absolute quantification*

A robust pipeline for using the water signal as internal reference for FID-MRSI should be established. To this end, the acquisition of unsuppressed water using the same sequence must be incorporated into the acquisition and quantification protocol. Additionally, a validation study should be performed by comparing the results with SVS in predefined brain regions, both in phantom and *in vivo* conditions. Depending on SVS voxel size, segmentation of the voxel into white and gray matter should be considered to ensure accurate and reliable quantification [67].

*Other applications*

The future of MRSI will likely involve implementing spectroscopic imaging versions of derived-MRS methods, such as dynamic MRS or diffusion-weighted MRS. In particular, there is growing interest in dynamic MRSI to monitor temporal metabolic changes *in vivo* with time resolutions below 5-10 minutes. This approach would be especially useful for studying brain energy metabolism through indirect *in vivo* detection of deuterated compounds [60].

Performing diffusion-weighted MRSI would also be highly valuable for simultaneously investigating different brain regions, where cellular microstructure varies significantly across areas [68]. Achieving high coverage and ensuring good data quality in the cerebellum could be particularly beneficial for studying developmental disorders.

Finally, alterations in macromolecular (MM) signals may serve as biomarkers for various diseases, and regional differences within the brain warrant further investigation in rodent models. However, the long acquisition times required to measure MM signals across multiple brain regions remains a major challenge. Mapping MM signals throughout the entire brain may become feasible with cryoprobes and 3D acquisitions, which offer significant potential for increasing SNR and accelerating data collection.

In summary, future advancements in MRSI will aim to enhance spatial coverage, acquisition efficiency, and reproducibility. These developments will pave the way for more comprehensive metabolic imaging, enabling deeper insights into brain function and pathology.

6. **Conclusion**

This article presents a comprehensive workflow for the acquisition and analysis of high-quality data using fast MRSI sequences (FID-MRSI) on preclinical scanners. By following the procedures described, novice users can reliably perform robust and accurate MRSI acquisitions at ultra-high magnetic fields (9.4T and 14.1T) in the rodent brain. Implementing the *MRS4Brain* toolbox or a custom pipeline that integrates all recommended steps will ensure reproducible and rigorous data analysis.

As a complement to this paper, we provide several educational videos demonstrating best practices for fast MRSI acquisition and processing via the *MRS4Brain* toolbox. Incorporating advanced strategies can further accelerate acquisition, improve spatial coverage, and enhance reproducibility. This paper provides a solid starting point for future developments in the field. Looking ahead, continued methodological developments will enable innovative metabolic imaging approaches, offering unprecedented insights into brain function and pathology.

**Appendix:**

**Point Spread Function**

Preclinical MRSI sequence optimization has to account for the Point Spread Function (PSF) and its influence on the spatial resolution and signal contamination. Indeed, in traditional phase-encoded MRSI, only 76.2% of the spectroscopic signal originates from the voxel acquired (66.7% in 3D) [69], with the rest coming from neighboring voxels (nominal voxel size increase due to large Full Width at Half Maximum (FWHM) of the central lobe) and distant voxels (non-canceled side lobes contributions). In $^1$H-MRSI, this can lead to an increase of lipid contamination if not taken care of during the acquisition, especially in the rodent brain where the lipid signal source, like ears and cheeks, are much closer to the region of interest.

The PSF can be altered by applying a different *k*-space sampling, either during the acquisition with specific encoding strategies [4, 70–72] or during the processing via a spatial apodization (or *k*-space filtering) [73, 74]. The principle of this alteration is to artificially improve signal quality by smoothing the side lobes of the PSF with minimal FWHM increase. The recommended filter is generally given by a Hamming function [74] as it allows for a large decrease of the lobe amplitude to below 1% with an increase of the nominal voxel size of a factor 1.8. The difference between the PSF with and without a Hamming filter applied can be observed in the work of Simicic et al. [10]. The Hamming function can also be achieved during the acquisition by using weighted averaging, with the correct averaging scheme to mimic a discretized Hamming profile at the cost of longer acquisition time [71].

## Acknowledgments

This work was supported by the Swiss National Science Foundation award no. 201218, 10000465, 10006046 and 207935 and by the CIBM Center for Biomedical Imaging of the UNIL, UNIGE, HUG, CHUV, EPFL. The development and refinement of these protocols represent a long-term collaborative effort within our group. We gratefully acknowledge the invaluable contributions of former PhD students (Dunja Simicic and Jessie Mosso) and we extend our sincere thanks to Wolfgang Bogner, Berhnard Strasser, to Ivan Tkac, and Vladimir Mlynarik for their expertise and support. The authors thank the veterinary staff at CIBM PCI EPFL for support during experiments. Their efforts were instrumental in overcoming technical challenges and ensuring the robustness and reproducibility of the protocols, which now serve as a cornerstone for our ongoing research.

## Authors' contribution

Study conception and design: GN, EM, CC

Acquisition of data: GN, BA, CC

Analysis and interpretation of data: GN, BA, CC

Drafting of manuscript: GN, EM, BA, AS, TTP, TNAD, TPL, BL, CC

Critical revision: GN, EM, BA, AS, TNAD, TPL, BL, CC

## Data Availability Statement

The *MRS4Brain toolbox* is available on the following repository: https://github.com/AlvBrayan/MRS4Brain-toolbox and MRS4Brain Toolbox – MRS4BRAIN - EPFL. Live demos on how to acquire *in vivo* [1]H-MRSI datasets, how to process *in vivo* acquired MM, how to use all the functionalities of the toolbox, as well as our scan protocols, basis sets, and example data sets are posted on the webpage: RESOURCES – MRS4BRAIN - EPFL.

**Supplemental Material**

**Supplemental Table 1: Minimum Reporting Standards in MRS for FID-MRSI at 9.4T and 14.1T**

**Hardware**

| | |
|---|---|
| **Field strength [T]** | 9.4T; 14.1T |
| **Manufacturer** | Bruker |

| Model (software) | 9.4T: Paravision 360 V3.5<br>14.1T: Paravision 360 V1.1 and V3.3 |
|---|---|
| RF coil | 9.4T: $^1$H-quadrature volume-transmit coil and a cryogenic four-channel receive rat head array<br>14.1T: $^1$H-quadrature surface rat head coil |
| Additional Hardware | N/A |

**Acquisition**

| Pulse Sequence | FID-MRSI (Bruker:CSI) |
|---|---|
| Volume of Interest (VOI) | Rodent: Brain |
| Nominal VOI Size | 0.77 × 0.77 × 2 mm$^3$ |
| Repetition Time (TR) and Acquisition Delay (AD) | 9.4T: TR = 813 ms / AD = 1.3 ms<br>14.1T: TR = 811.48 ms / AD = 1.3 ms |
| Number of Averaged Spectra (NA) | 1 average |
| FOV in All Directions | 24 × 24 × 2 mm$^3$ FOV |
| Matrix Size | 31×31 |
| Spectral Bandwidth (Hz) | 9.4T: 5000<br>14.1T: 7142.85 |
| Number of Spectral Points | 9.4T: 768<br>14.1T: 1024 |
| Water Suppression Method | VAPOR |
| Shimming Method | Bruker MAPSHIM, first in an ellipsoid covering the full brain and further in a volume of interest centered on the MRSI slice, with a thickness of 2 mm |
| Triggering or Motion Correction | N/A |

**Data Analysis**

| Analysis Software | LCmodel (Version 6.3-1N) |
|---|---|
| Processing Step Parameters | Custom Basis-Sets for AD = 1.3ms (for both 9.4T and 14.1T)<br>Control files provided with the *MRS4Brain Toolbox* |
| Output Measure | Ratios to total Creatine |
| Quantification Reference | Basis-set including: alanine, aspartate, ascorbate, creatine, phosphocreatine, γ-aminobutyrate, glutamine, glutamate, glycerophosphocholine, glutathione, glucose, inositol, N-acetylaspartate, N-acetylaspartylglutamate, phosphocholine, phosphoethanolamine, lactate, taurine simulated using NMR ScopeB. Macromolecules acquired in-vivo with double inversion recovery FID-MRSI. |

### Data quality

| | |
|---|---|
| **Reported Variables** | SNR (reference to NAA) and linewidths (reference to water) both reported<br>Global linewidths estimated by LCModel |
| **Data Exclusion Criteria** | LCModel SNR > 4 and<br>LCModel FWHM < 1.25*LCModel |
| **Quality Measures of Post-processing Model Fitting** | CRLB < 30% |
| **Sample Spectrum** | Figure 6 |

## Supplemental Table 2: MRSinMRS for PRESS-MRSI

### Hardware

| | |
|---|---|
| **Field Strength [T]** | 14.1 |
| **Manufacturer** | Bruker |
| **Model (software)** | Paravision 360 V3.3 |
| **RF coil** | $^1$H-quadrature surface rat head coil |
| **Additional Hardware** | N/A |

### Acquisition

| | |
|---|---|
| **Pulse Sequence** | PRESS-MRSI (Bruker:CSI) |
| **Volume of Interest (VOI)** | Rodent: Brain |
| **Nominal VOI Size** | 0.77 × 0.77 × 2 mm$^3$ |
| **Repetition Time (TR) and Echo Time (TE)** | TR = 2000 ms / TE = 10.2 ms |
| **Number of Averaged Spectra (NA)** | 1 average |
| **FOV in All Directions** | 24 × 24 × 2 mm$^3$ |
| **Matrix Size** | 31×31 |
| **Spectral Bandwidth (Hz)** | 7142.85 |
| **Number of Spectral Points** | 1024 |
| **Water Suppression Method** | VAPOR |
| **Shimming Method** | Bruker MAPSHIM, first in an ellipsoid covering the full brain and further in a volume of interest centered on the MRSI slice, with a thickness of 2 mm |
| **Triggering or Motion Correction** | N/A |

### Data Analysis

| | |
|---|---|
| **Analysis Software** | LCmodel (Version 6.3-1N) |

| Processing Step Parameters | Custom Basis-Set for TE = 10.2 ms<br>Control files provided with the *MRS4Brain Toolbox* |
|---|---|
| Output Measure | Ratios to total Creatine |
| Quantification Reference | Basis-set including: alanine, aspartate, ascorbate, creatine, phosphocreatine, γ-aminobutyrate, glutamine, glutamate, glycerophosphocholine, glutathione, glucose, inositol, N-acetylaspartate, N-acetylaspartylglutamate, phosphocholine, phosphoethanolamine, lactate, taurine simulated using NMR ScopeB. Macromolecules acquired in-vivo with double inversion recovery FID-MRSI. |

**Data quality**

| Reported Variables | SNR (reference to NAA) and linewidths (reference to water) both reported<br>Global linewidths estimated by LCModel |
|---|---|
| Data Exclusion Criteria | LCModel SNR > 4 and<br>LCModel FWHM < 1.25*LCModel |
| Quality Measures of Post-processing Model Fitting | CRLB < 30% |
| Sample Spectrum | Figure 6 |

**Supplemental Table 3: Minimum Reporting Standards in MRS for SE-MRSI**

**Hardware**

| Field strength [T] | 9.4T |
|---|---|
| Manufacturer | Bruker |
| Model (software) | Paravision 360 V3.5 |
| RF coil | $^1$H-quadrature volume-transmit coil and a cryogenic four-channel receive rat head array |
| Additional Hardware | N/A |

**Acquisition**

| Pulse Sequence | SE-MRSI (Bruker:CSI) |
|---|---|
| Volume of Interest (VOI) | Rodent: Brain |
| Nominal VOI Size | 0.77 × 0.77 × 2 mm$^3$ |
| Repetition Time (TR) and Echo Time (TE) | TR = 2000 ms / TE = 3.284 ms |
| Number of Averaged Spectra (NA) | 1 average |
| **FOV in All Directions** | 24 × 24 × 2 mm$^3$ |
| **Matrix Size** | 31×31 |
| **Spectral Bandwidth (Hz)** | 5000 |
| **Number of Spectral Points** | 768 |
| **Water Suppression Method** | VAPOR |
| **Shimming Method** | Bruker MAPSHIM, first in an ellipsoid covering the full brain and further in a volume of interest centered on the MRSI slice, with a thickness of 2 mm |
| **Triggering or Motion Correction** | N/A |

**Data Analysis**

| | |
|---|---|
| **Analysis Software** | LCmodel (Version 6.3-1N) |
| **Processing Step Parameters** | Custom Basis-Sets for STEAM TE = 3 ms<br>Control files provided with the *MRS4Brain Toolbox* |
| **Output Measure** | Ratios to total Creatine |
| **Quantification reference** | Basis-set including: alanine, aspartate, ascorbate, creatine, phosphocreatine, γ-aminobutyrate, glutamine, glutamate, glycerophosphocholine, glutathione, glucose, inositol, N-acetylaspartate, N-acetylaspartylglutamate, phosphocholine, phosphoethanolamine, lactate, taurine simulated using NMR ScopeB. Macromolecules acquired in-vivo with double inversion recovery FID-MRSI. |

**Data quality**

| | |
|---|---|
| **Reported Variables** | SNR (reference to NAA) and linewidths (reference to water) both reported<br>Global linewidths estimated by LCModel |
| **Data Exclusion Criteria** | LCModel SNR > 4 and<br>LCModel FWHM < 1.25*LCModel |
| **Quality Measures of Post-processing Model Fitting** | CRLB < 30% |
| **Sample Spectrum** | Figure 6 |

**Supplemental Table 4: MRSinMRS for SVS STEAM Metabolites and Water acquisitions used for quality assessment.**

**Hardware**

| | |
|---|---|
| **Field strength [T]** | 9.4T |
| **Manufacturer** | Bruker |
| **Model (software)** | 9.4T: Paravision 360 V3.5 |
| **RF coil** | [1]H-quadrature volume-transmit coil and a cryogenic four-channel receive rat head array |
| **Additional Hardware** | N/A |

### Acquisition

| | |
|---|---|
| **Pulse Sequence** | Bruker: STEAM |
| **Volume of Interest (VOI)** | Rodent: Brain |
| **Nominal VOI Size** | 10 × 10 × 2 mm$^3$ |
| **Repetition Time (TR), Echo Time (TE), Mixing Time (TM)** | TR = 4000 ms / TE = 3 ms / TM = 10 ms |
| **Number of Averaged Spectra (NA)** | 32 (metabolites); 16 (water) |
| **FOV in All Directions** | NA |
| **Matrix Size** | NA |
| **Spectral Bandwidth (Hz)** | 5000 |
| **Number of Spectral Points** | 4096 |
| **Water Suppression Method** | VAPOR (metabolites); NO SUPPRESSION (water) |
| **Shimming Method** | Bruker MAPSHIM, first in an ellipsoid covering the full brain and further in a volume of interest centered on the voxel |
| **Triggering or Motion Correction** | N/A |

### Data Analysis

| | |
|---|---|
| **Analysis Software** | LCmodel (Version 6.3-1N) |
| **Processing step Parameters** | Custom Basis-Sets for STEAM TE = 3 ms<br>Control files provided with the *MRS4Brain Toolbox* |
| **Output measure** | Ratios to total Creatine |
| **Quantification reference** | Basis-set including: alanine, aspartate, ascorbate, creatine, phosphocreatine, γ-aminobutyrate, glutamine, glutamate, glycerophosphocholine, glutathione, glucose, inositol, N-acetylaspartate, N-acetylaspartylglutamate, phosphocholine, phosphoethanolamine, lactate, taurine simulated using NMR ScopeB. Macromolecules acquired in-vivo with double inversion recovery FID-MRSI. |

### Data quality

| | |
|---|---|
| **Reported Variables** | SNR (reference to NAA) and linewidths (reference to water) both reported<br>Global linewidths estimated by LCModel |
| **Data Exclusion Criteria** | LCModel SNR > 4 and |
| **Quality Measures of Post-processing Model Fitting** | LCModel FWHM < 1.25*LCModel<br>CRLB < 30% |
| **Sample Spectrum** | Figure 6 |